\documentclass[a4paper,12pt]{article}

\usepackage[T2A]{fontenc}
\usepackage{amsmath,amssymb}
\usepackage{cite}
\usepackage[linktocpage=true,plainpages=false,pdfpagelabels=false]{hyperref}

\usepackage[utf8]{inputenc}
\usepackage[english]{babel}

\usepackage{pgfplots}

\begin{document}

\title{Classification of ten-dimensional symmetric embeddings for the spatially flat Friedman model}

\author{
M.~S.~Sushkov\thanks{E-mail: matveysushkov34@gmail.com},
S.~A.~Paston\thanks{E-mail: pastonsergey@gmail.com}\\
{\it Saint Petersburg State University, Saint Petersburg, Russia}
}
\date{\vskip 15mm}
\maketitle

\begin{abstract}
All possible variants of symmetric embedding of the metric of the spatially flat Friedman model into a ten-dimensional ambient space are analyzed. It is shown that only two such embeddings exist: the five-dimensional embedding found by Robertson in 1933 and a new eight-dimensional embedding proposed in this work.
The new embedding can be utilized in developing the idea of explaining dark matter as fictitious matter arising in the description of gravity within the embedding theory approach. Some results in this direction are presented. Additionally, the possibility of constructing embeddings of the spatially flat Friedman model whose symmetry is lower than that of the metric itself is discussed. Two such embeddings are constructed.
\end{abstract}

\newpage

\section{Introduction}
Embedding theory is one of the possible modifications of gravity. This theory represents curved spacetime as a four-dimensional surface embedded in an $N$-dimensional ambient space. This approach to describing gravity was proposed in 1975 by T. Regge and C. Teitelboim and published in \cite{regge}. Within the framework of embedding theory, the surface corresponding to spacetime is defined using embedding functions $y^a(x^\mu)$, and it has an induced metric:
\begin{equation}\label{induced}
g_{\mu \nu}= \partial_{\mu}y^{a} \partial_{\nu}y^{b} \eta_{ab},
\end{equation}
where $\eta_{ab}$ is the flat metric of the ambient space. Such a modification of gravity is similar to mimetic gravity \cite{mukhanov, Golovnev201439} but has a clearer geometric meaning \cite{statja51}. Both of these theories belong to the class of gravity theories obtained as a result of differential transformations of field variables \cite{statja60}.

The equations of motion in embedding theory, also known as the Regge-Teitelboim equations, can be written using the second fundamental form of the surface $b_{\mu \nu} = D_\mu D_\nu y^a$:
\begin{equation}\label{RegTei}
(G^{\mu \nu} - \varkappa T^{\mu \nu})b_{\mu \nu}^{a} = 0,
\end{equation}
where $D_\mu$ denotes the covariant derivative. Clearly, the solutions to these equations include not only the solutions of Einstein's equations $G^{\mu \nu} = \varkappa T^{\mu \nu}$ but also additional solutions. As noted in \cite{pavsic85}, the equations \eqref{RegTei} can be rewritten as:
\begin{equation}\label{bigRT}
\begin{aligned}
    &G^{\mu \nu} = \varkappa (T^{\mu \nu} + \tau^{\mu \nu}),\\
    &\tau^{\mu \nu}b_{\mu \nu}^{a} =0.
\end{aligned}
\end{equation}
Thus, the equations of motion in embedding theory are equivalent to Einstein's equations with additional matter that obeys certain equations of motion. Here, $\tau^{\mu\nu}$ acts as the energy-momentum tensor of this matter, which can be referred to as embedding matter.

This matter affects the gravitational field but does not possess any evident means of interacting with ordinary matter except through gravity. Given that experiments to detect dark matter through non-gravitational means have not yet been successful \cite{1509.08767, 1604.00014}, embedding matter can be considered as a potential candidate for dark matter.

In the equations \eqref{induced} and \eqref{bigRT}, the unknown variables are $g^{\mu \nu}, \tau^{\mu\nu}, y^a$. To analyze the properties of embedding matter and compare them with the known characteristics of dark matter, it is necessary to find the tensor $\tau^{\mu \nu}$ that satisfies these equations. The concept of dark matter is closely related to cosmological models characterized by high symmetry. When studying the properties of embedding matter within cosmological models, an ansatz for the unknown variables describing it can be obtained. In this context, the left-hand side of \eqref{induced} corresponds to the Friedman metric with an arbitrary scale factor $a(t)$. Moreover, the tensor $\tau^{\mu \nu}$ in these cosmological models becomes diagonal and can be expressed through two time-dependent functions. Formulating the conditions imposed by symmetry on the variable $y^a$ is considerably more complex. The method for imposing such conditions is described in Section 3. As is currently known \cite{WMAP9}, the size of the observable universe is significantly smaller than its curvature radius. For this reason, we will consider the spatially flat Friedman model as the spacetime.

According to the Janet-Cartan-Friedman theorem \cite{fridman61}, a curved space of dimension $n$ can always be locally embedded into an ambient space of dimension ${n(n+1)}/2$. Thus, for at least a local embedding of the spatially flat Friedman model, ten dimensions of the ambient space are sufficient. Therefore, we will embed this model into $\mathbb{R}^{1,9}$.

The spatially flat Friedman model is symmetric under spatial translations and rotations. Naturally, it can be assumed that the corresponding surface in the ambient space will also possess this symmetry. Using the method proposed in \cite{statja27}, we can take this fact into account, significantly simplifying the nonlinear partial differential equations for the embedding functions \eqref{induced}. The aim of this work is to find all possible explicit forms of symmetric embeddings of the spatially flat Friedman model into the flat ten-dimensional ambient space $\mathbb{R}^{1,9}$. A similar problem for a five-dimensional ambient space was solved in \cite{statja29}, where it was shown that the only solution is the embedding first proposed in \cite{robertson1933}.

Section 2 discusses the concept of embedding unfoldedness and justifies the relevance of this property for embeddings. Section 3 presents the method for constructing a surface symmetric under Friedman symmetries, as well as describing two previously unknown embeddings with reduced symmetry. Section 4 provides a classification of all symmetric embeddings of the spatially flat Friedman model into $\mathbb{R}^{1,9}$. Section 5 examines in detail the new symmetric eight-dimensional embedding and its properties.

\section{Unfolded embeddings}
The concept of the unfoldedness of a surface in an $N$-dimensional embedding space was discussed in \cite{statja71}. To define this concept, it is necessary to introduce the second
fundamental form of the surface:
\begin{equation}
b_{\mu \nu}^{a} = D_{\mu}D_{\nu} y^{a} = D_\mu \partial_\nu y^a.
\end{equation}
It is straightforward to prove that this quantity is orthogonal to the surface at every point:
\begin{equation}\label{Proector} b_{\mu \nu}^{a} = \partial_{\mu}\partial_{\nu}y^{a} - \Gamma^{\alpha}_{\mu\nu} \partial_{\alpha}y^{a} = \partial_{\mu} \partial_{\nu}y^{a} - \partial_{\alpha}y^{a} (\partial_{\mu}\partial_{\nu}y^{b} \partial^{\alpha} y^{c})\eta_{bc} = \Pi_{b \perp}^{a}\partial_{\mu}\partial_{\nu}y^{b}, \end{equation}
where $\Pi_{b\perp}^{a}(x)$ is the projector onto the subspace orthogonal to the surface at the point $x$, and $\partial^{\alpha} = g^{\alpha \beta} \partial_\beta$. The equality $\Gamma^\alpha_{\mu \nu} = (\partial_{\mu}\partial_{\nu}y^{b} \partial^{\alpha} y^{c})\eta_{bc}$ used in the formula above follows from the expression for the connection in terms of the metric and the metric in terms of the embedding functions from \eqref{induced}.

After introducing a basis in the subspace orthogonal to the surface at each point, the index $a$ in the tensor $b^a_{\mu \nu}$ can be replaced with an index that runs over $N-4$ values. Using the symmetry of the tensor $b^a_{\mu \nu}$, the indices $\mu \nu$ can be replaced with a multi-index running over 10 values. Thus, the tensor $b^a_{\mu \nu}$ can be uniquely represented as a matrix of size $(N-4) \times 10$. The unfoldedness of the surface implies the maximal non-degeneracy of the tensor $b_{\mu\nu}^a$, meaning that the rank of the corresponding $(N-4) \times 10$-matrix is maximal at almost all points of the surface.

Unfoldedness of the surface leads to several consequences. The first is that the neighborhood of almost any point is contained (up to the second order) within all $N$ dimensions for $N \leq 14$. Near a point $q^\mu$ on the surface, the embedding functions can be expanded in a Taylor series:
\begin{align} &y^a(q^\mu +x^\mu) = y^a(q^\mu) + \partial_\nu y^a(q^\mu) x^\nu + \partial_\nu \partial_\gamma y^a(q^\mu)x^\nu x^\gamma + ... = \nonumber\\
&=y^a(q^\mu) + \partial_\nu y^a(q^\mu) x^\nu + \Pi_b^a\partial_\nu \partial_\gamma y^b(q^\mu)x^\nu x^\gamma + \Pi_{b\perp}^a\partial_\nu \partial_\gamma y^b(q^\mu)x^\nu x^\gamma + ... = \nonumber\\
&=y^a(q^\mu) + \partial_\nu y^a(q^\mu) x^\nu + \Pi_b^a\partial_\nu \partial_\gamma y^b(q^\mu)x^\nu x^\gamma + b^a_{\nu \gamma}(q^\mu)x^\nu x^\gamma + ... 
\end{align}
In the last equality, we used \eqref{Proector}. Here, $\Pi_{b}^{a}(q)$ denotes the projector onto the tangent subspace at the point $q$. The terms $\partial_\nu y^a(q^\mu) x^\nu$ and $\Pi_b^a\partial_\nu \partial_\gamma y^b(q^\mu)x^\nu x^\gamma$ lie entirely within the 4-dimensional tangent subspace of the surface. The term $b^a_{\nu \gamma}(q^\mu)x^\nu x^\gamma$ lies entirely within the orthogonal subspace, the dimensionality of which coincides with the rank of the matrix $b^a_{\nu \gamma}$. If its rank is less than $N-4$ at a certain point, then, up to the second order w.r.t. $x^\mu$, the surface in the vicinity of that point will lie in a subspace of dimensionality less than $N$, meaning it can be further "unfolded" \  into untouched dimensions. This is why the condition on the tensor $b_{\mu \nu}^{a}$ described above is referred to as the unfoldedness of the surface.

The second consequence is the relationship between small metric deformations and small surface deformations. For $N=10$, this relationship is bijective for an unfolded surface, which leads to the superposition principle for weak gravitational fields if the embedding of the background metric is unfolded. Details can be found in \cite{2311.02515}. For $N=14$, the deformation of the metric is bijectively related to the transverse deformation of the surface. The case $N < 10$ remains largely unexplored and requires further investigation, so it is currently difficult to assess the role of the unfoldedness concept in this scenario.

Let us describe this relationship. The embedded surface possesses the induced metric \eqref{induced}. Let us perform a small deformation of the surface: $y^{a} \rightarrow y^{a} + \delta y^{a}$. Decompose $\delta y$ into tangential and orthogonal parts: $\delta y^{a} = \xi^{\mu}\partial_{\mu}y^{a} + \delta y^{a}_{\perp}$. A deformation of the surface will lead to a first-order change in the metric:
\begin{equation}\label{deltag} \begin{aligned} & \delta g_{\mu \nu} = ( \partial_{\mu} (\xi^{\gamma}\partial_{\gamma}y + \delta y_{\perp}), \partial_{\nu}y) + (\partial_{\mu} y, \partial_{\nu}(\xi^{\gamma}\partial_{\gamma}y + \delta y_{\perp})) = \\
& = g_{\gamma \nu} \partial_{\mu}\xi^{\gamma} + g_{\mu \gamma} \partial_{\nu}\xi^{\gamma} + \xi^{\gamma}(\Gamma_{\mu \gamma}^{\alpha}g_{\alpha \nu} + \Gamma_{\nu \gamma}^{\alpha}g_{\alpha \mu}) - \\
& - 2(\partial_{\mu} \partial_{\nu} y , \delta y_{\perp}) = \\
& = D_{\mu}\xi_{\nu} + D_{\nu}\xi_{\mu} - 2 b_{\mu \nu}^{a} \delta y_{\perp a}. \end{aligned} \end{equation}
The contributions corresponding to the longitudinal deformation of the surface, $D_{\mu}\xi_{\nu} + D_{\nu}\xi_{\mu}$, can be eliminated using the general covariance of the metric. The remaining term, $- 2 b_{\mu \nu}^{a} \delta y_{\perp a}$, is therefore the deformation of the metric that cannot be reduced to a diffeomorphism.


\section{Symmetric embeddings}
To explicitly construct embeddings of a space with a given metric, it is necessary to solve the equations for the induced metric \eqref{induced}. These equations are nonlinear first-order partial differential equations. Finding all solutions to these equations is often infeasible, even when the embedding space has a relatively small number of dimensions. Therefore, it is reasonable to employ variable separation techniques prior to the solving.

One possible approach to achieve this separation is to account for the symmetry of the space whose embedding functions need to be determined. In this case, it seems natural to require that the image of a symmetric manifold retains its symmetry. A method for incorporating symmetry into the construction of embeddings was proposed in \cite{statja27}.

Let us denote the embedded surface as $M$. Suppose the embedding functions map a 4-dimensional pseudo-Riemannian space $L$, described by coordinates $x$, into the surface $M$. Let the group $G$ act on $L$, and let $L$ be symmetric with respect to this group, meaning that distances between nearby points remain unchanged under the group's action. It is said that the surface $M$ is symmetric with respect to the group $G$ if it is invariant under the action of a subgroup (which is isomorphic to the group $G$) of the group of motions $\mathcal{P}$ of the flat embedding space. To construct the desired surface in this manner, it is necessary to find a homomorphism $V$ that maps the group $G$ into the group $\mathcal{P}$. Consequently, $V$ is a representation of the group $G$. This representation must be unique and exact. Having the representation $V$ of the group $G$, we can act with it on a given initial vector $\tilde y$ in the embedding space. Then, $M = V(G) \tilde y$ will be a surface in the embedding space that is guaranteed to be symmetric with respect to the group $G$.

The remaining task is to choose the initial vector $\tilde y$. It may happen that the surface $V(G) \tilde y$ has a dimensionality different from that of the original space $L$, which is unacceptable. To select a suitable initial vector, we assume that the desired embedding functions, determined by the representation $V(g)$, map the points of the manifold $L$ to the points of the surface $M$ in a one-to-one manner. Therefore, if for a point $\tilde x \in L$, corresponding to the initial vector $\tilde y$, there exists a stabilizer subgroup $H$ in the group $G$, such that $H \tilde x = \tilde x$, then the image of this subgroup $V(H)$ must also leave the initial vector $\tilde y$ invariant:
$V(H) \tilde y = \tilde y.$

\subsection{Example: $SO(3)$ symmetry}
From now on, we will consider the spatially flat Friedmann model as the base space $L$. This space is symmetric under the action of the group $SO(3) \ltimes T(3)$ (where $T(3)$ denotes the three-dimensional translation group) and its metric in spherical coordinates is given by:
\begin{equation}\label{metricSphere}
    g_{\mu \nu } =
    \begin{pmatrix}
        1 & 0 & 0 & 0 \\
        0 & -a^2(t) & 0& 0 \\
        0 & 0 & -a^2 (t)r^2 & 0 \\
        0 & 0 & 0 & -a^2(t)r^2\sin^2 \theta
    \end{pmatrix}.
\end{equation}

As an example, let us consider the partial symmetry of this space with $G = SO(3)$. Let us take $\mathbb{R}^{1,9}$ as the embedding space. One possible representation of the group $G$ can be written as follows: \begin{equation}
    V(O) =
    \begin{pmatrix}
        1 & 0 & 0 & 0 \\
        0 & O_{ab} & 0& 0 \\
        0 & 0 &  \lambda^A_{ij}\lambda^B_{kl}O_{ik}O_{jl} & 0 \\
        0 & 0 & 0 & 1
    \end{pmatrix},
\end{equation}
where $O \in SO(3)$ is an orthogonal matrix, and $\lambda^A_{ij}$ is a basis in a five-dimensional space of traceless symmetric second-rank tensors normalized as follows: $\lambda^A_{ij} \lambda^B_{ij} = \delta^{AB}$. It can be observed that the first and last blocks of $V(O)$ correspond to the scalar representation of $SO(3)$, the block $O_{ab}$ corresponds to the vector representation, and $\lambda^A \lambda^B OO$ corresponds to the five-dimensional irreducible representation. This embedding was obtained in \cite{statja76}.

Since in the space $L$, the symmetry $G = SO(3)$ leaves the coordinates $r$ and $t$ unchanged, the initial vector must depend on them. One possible choice for the initial vector is:
\begin{equation}
    \tilde y  =
    \begin{pmatrix}
        w(t,r) \\ f(t,r) e_b \\ u(t,r) \lambda^B_{ij}e_i e_j \\ h(t,r)
    \end{pmatrix}, \ \
e_a =
\begin{pmatrix}
    0 \\ 0 \\ 1
\end{pmatrix}.
\end{equation}
Then the representation $V(O)$, acting on the initial vector, gives a surface with the following parametrization: \begin{equation}\label{anzazSO3}
\begin{aligned}
&y^{0} = w(t,r), \\
&y^{i} = f(t,r) n^i, \\
&y^{A} = u(t,r) \lambda^A_{ik}n^in^k, \ A = 4,...,8,\\
&y^{9} = h(t,r),
\end{aligned}
\end{equation}
where $n_i (\theta, \varphi) = O_{ij}e_j$ is a unit vector.

Thus, we have obtained an ansatz for solving the equation for the induced metric \eqref{induced}. Using this ansatz, an explicit embedding of the spatially flat Friedmann metric into the flat embedding space $\mathbb{R}^{1,9}$ with the metric $\eta_{ab} = diag(1,-1,-1,...,-1)$ was obtained: \begin{equation}\label{Emb_19_SO3}
\begin{aligned}
& y^{0} = a(t) \frac{r^{2}\left(1-\frac{2}{3}\sin^{2}\gamma(r)\right)}{2c_{1}} + \frac{c_{1}a(t)}{2}+ \int\frac{1}{2c_{1}\dot{a}(t)}dt, \\
& y^{i} = a(t)r n^i \cos \gamma(r), \\
& y^{A} = a(t)r \lambda^A_{ik}n^in^k \sin \gamma(r), \\
& y^{9} = a(t) \frac{r^{2}\left(1-\frac{2}{3}\sin^{2}\gamma(r)\right)}{2c_{1}} - \frac{c_{1}a(t)}{2}+ \int\frac{1}{2c_{1}\dot{a}(t)}dt,
\end{aligned}
\end{equation}
where $\gamma(r)$ is implicitly defined by the equation \begin{equation}\label{alpha}
r = \widetilde{c} \left(\frac{1\pm \cos \gamma(r)}{|\sin \gamma(r)|}\right)^{\sqrt{\frac{3}{2}}} \frac{1}{|\sin \gamma(r)|}.
\end{equation}
The function $\gamma(r)$ has the asymptotic behavior $\gamma(r) \sim r^{2+\sqrt{6}}$ as $r \rightarrow 0$. Consequently, the embedding functions $y^A \sim r \sin\gamma(r)$ exhibit the asymptotic behavior $y^A \sim r^{3+\sqrt{6}}$ near $r = 0$, which ensures that they are five times differentiable at this point.

\begin{figure}[htbp]
    \centering
    \includegraphics[width=0.6\textwidth]{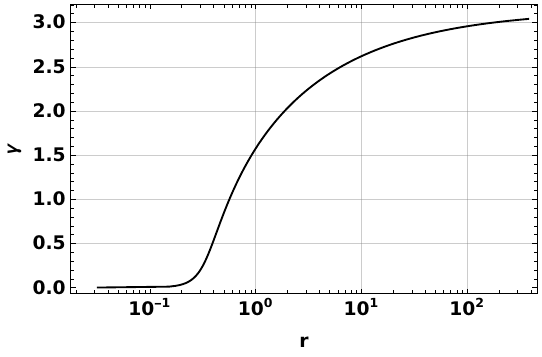} 
    \caption{Dependence of $\gamma(r)$ on $r$.} 
    \label{fig:graph} 
\end{figure}

The embedding functions \eqref{Emb_19_SO3}, along with the function $\gamma(r)$ satisfying equation \eqref{alpha}, solve the system of equations for the induced metric. The function $\gamma(r)$ monotonically increases from $0$ to $\pi$ as $r$ increases from $0$ to $\infty$. Its graph is presented in Fig. \ref{fig:graph}.
The surface corresponding to this embedding turned out to be unfolded. Moreover, it is infinitely differentiable everywhere except at the point $r=0$, where it has only five derivatives.

The ansatz \eqref{anzazSO3} is also suitable for finding $SO(3)$-symmetric embeddings in the space $\mathbb{R}^{2,8}$ with the embedding space metric $\eta_{ab} = diag(1,-1,...,-1,1)$. In this case, an explicit embedding expressed in terms of Jacobi elliptic functions was obtained: \begin{equation}\label{Emb_28}
\begin{aligned}
& y^{0} = a(t) r \ dn\left(\sqrt{3}\ln\left(\frac {r}{r_0}\right), \frac23  \right), \\
& y^{i} = a(t)r\ cn\left(\sqrt{3}\ln\left(\frac {r}{r_0}\right), \frac23  \right) n^i , \\
& y^{A} = a(t)r\ sn\left(\sqrt{3}\ln\left(\frac {r}{r_0}\right), \frac23  \right)  \lambda^A_{ik}n^in^k,\\
& y^{9} = t,
\end{aligned}
\end{equation} where $r_0$ is an arbitrary constant, $sn, cn$ are the Jacobi elliptic sine and cosine functions, and $dn$ is the Jacobi delta amplitude. This solution is also unfolded. It is infinitely differentiable at all points except $r=0$, where it is only continuous.

\subsection{Full symmetry $SO(3)\ltimes T(3)$}\label{fullsym}
Let us now take into account the full symmetry of the spatially flat Friedmann model. The symmetry group is $G = SO(3) \ltimes T(3)$. The action of this group on a vector from the Friedmann space $x = \begin{pmatrix}
    t \\ x_i
\end{pmatrix}$ is understood as $g x = \begin{pmatrix}
    t \\ O_{ij}x_j + a_i
\end{pmatrix}$.

In 1933, Robertson obtained an embedding in a 5-dimensional space symmetric with respect to the group $G$ \cite{robertson1933}: \begin{equation}\label{robertson}
    \begin{aligned}
        & y^{+} = \frac{1}{2} \left(r^{2} a(t) + \int \frac{dt}{\dot{a}(t)}\right),\\
        & y^{-} = a(t),\\
        & y^{2} = a(t) r \cos \theta ,\\
        & y^{3} = a(t) r \sin \theta \cos \varphi,\\
        & y^{4} = a(t) r \sin \theta  \sin \varphi.
    \end{aligned}
\end{equation}
The indices $+$ and $-$ correspond to lightlike coordinates in the ambient space. This solution, as an embedding of a 4-dimensional space into a 5-dimensional one, is unfolded.

We aim to find all symmetric embeddings of the metric \eqref{metricSphere} of the spatially flat Friedmann model into $\mathbb{R}^{1,9}$. The transformation group of this space is $\mathcal{P}(\mathbb{R}^{1,9}) = SO(1,9)\ltimes T(10)$. We will use the standard notation for the action of a rotation $\Omega \in SO(1,9)$ and a translation by a vector $\vec{b} \in \mathbb{R}^{1,9}$ on a point in the ambient space $\vec{y}$: \begin{equation} \label{calP}
    (\Omega, \vec{b}) \circ \vec{y} =
    \begin{pmatrix}
        \Omega & \vec{b} \\
        0 & 1
    \end{pmatrix}
    \begin{pmatrix}
        \vec{y} \\ 1
    \end{pmatrix}.
\end{equation}
The matrix $V(O \times a)$ must take the form of the matrix on the right-hand side of \eqref{calP}.

The group $G$ is a semidirect product of the group $SO(3)$ and the translation group $T(3)$. This means that its representations can be written as: \begin{equation}\label{repr decomp}
    V(O \times a) = e^{a_i p_i}V(O \times 0),
\end{equation}
where $O$ is an element of the group $SO(3)$, $a$ is from the group $T(3)$, and $p_i$ are mutually commuting matrices of the same size as $V$, satisfying the condition
\begin{equation}\label{p universal}
    V(O)p_i V^{-1}(O) = O_{ik} p_k,
\end{equation}
where $V(O) = V(O \times 0)$.

It is evident that if we have constructed a surface, representing the same surface in a new basis will not yield a new result. For this reason, we can start the construction of the embedding with a basis in which $V(O)$ becomes block-diagonal with irreducible blocks along the diagonal. The formula \eqref{p universal} in this basis can be rewritten for individual blocks: \begin{equation} V_{j_{1}j_{2}} p_{i j_2 l_1} V^{-1}_{l_1 l_2} = O_{ik} p_{k j_1 l_2}, \end{equation} showing that in this basis, the blocks of the matrix $p_i$ are proportional to universal tensors of appropriate dimensions, i.e., tensors that remain invariant under the group action. For the group $SO(3)$, all such tensors can be explicitly written as: \begin{equation}\label{universal} \delta_i^k, \varepsilon_{ijk}, \lambda^A_{ij}, \lambda^A_{ijk}, \lambda^A_{ijkl}. \end{equation} Here, $\lambda_{ij}^A$ is a basis in the 5-dimensional space of symmetric traceless tensors with two $SO(3)$ indices, $\lambda_{ijk}^A$ is in the 7-dimensional space with three indices, and $\lambda_{ijkl}^A$ is in the 9-dimensional space with four indices. The index $A$ runs through all elements of the respective basis, and the group $SO(3)$ acts on index $A$ in its corresponding representation. Below, we explain why we do not include bases corresponding to representations with half-integer spin in this list.

Note that in the basis where the matrix $V(O)$ has a block-diagonal form, there must be at least two scalar blocks among the irreducible blocks. One of these blocks corresponds to the 11th component, which ensures that translations in the ambient space are implemented in \eqref{calP}, while the other scalar block corresponds to time in the ambient space.

Since the symmetry group of the space $L$ leaves only the coordinate $t$ invariant, the initial vector will be a function of time only. As the dependence on the variables $r, \theta, \varphi$ is fully determined by the representation matrix $V$ when constructing the symmetric surface, the PDE \eqref{induced} can be reduced to ODEs for the initial vector $\tilde y (t)$.

Note that any initial vector
$\Tilde{x}(t) = \begin{pmatrix}
    t \\ \Tilde{\mathbf{x}}_{i}(t)
\end{pmatrix}$ generates a stability subgroup isomorphic to the group $SO(3)$ --- rotations around the initial point $\Tilde{\mathbf{x}}_i(t)$. However, considering that the initial vector can always be set to zero through a change of coordinates, the stabilizer subgroup can be chosen as group $SO(3)$ --- rotations around the origin. In this case, the initial vector $\tilde y(t)$ should be chosen in such way that the matrix $V(SO(3))$ leaves it invariant. Consequently, the surface generated by the action of $V(O \times a)$ on $\tilde y(t)$ will be 3-dimensional for each value of $t$, and thus, overall, 4-dimensional. Indeed, the algebra of the group $SO(3) \ltimes T(3)$ consists of 6 generators, 3 of which belong to $SO(3)$ and generate the stabilizer subgroup, meaning they do not change the dimensionality of the surface.The action of the remaining 3 generators, namely the translation generators in 3-dimensional space, will result in a 3-dimensional surface in the ambient space for each value of $t$.

Thus, the representation $V(O \times a)$ must be a one-to-one homomorphism. If the action of the group $SO(3)$ on a vector in the original space $L$ yields a 0-dimensional point for the origin or a 2-dimensional sphere for other cases, the image of this surface under the action of the representation $V$ must have the same dimensionality. Suppose that among the irreducible blocks of the representation $V$ there is a spinor representation of $SO(3)$. The subspace corresponding to this representation is called the spinor subspace. Then, if at some point on the surface the coordinates in the spinor subspace are nonzero, the action of $V(O)$ on this point will produce a 3-dimensional surface, not a 2-dimensional or 0-dimensional one. For this reason, we do not consider spinor representations among the irreducible ones.

In conclusion, to construct symmetric embedding functions mapping the spatially flat Friedmann model into $\mathbb{R}^{1,9}$, the following steps must be taken. \begin{enumerate}
    \item Choose a set of irreducible tensor representations of the group $SO(3)$ whose total dimensions sum to 11. This set must include at least two scalar irreducible representations. \item Fill the matrix $p_i$ with blocks of appropriate sizes proportional to the universal tensors \eqref{universal} of the group $SO(3)$.
    \item Adjust the coefficients of the blocks in $p_i$ to ensure that the matrices $p_i$ commute with each other.
    \item Obtain the representation of the symmetry group using formula \eqref{repr decomp}.
    \item Adjust the coefficients of the $p_i$ matrices to bring the matrix $V(O \times a)$ into the form \eqref{calP}. \item Find a vector $\tilde y(t)$ such that $V(SO(3))\tilde y(t) = \tilde y(t)$. The action of the obtained matrix $V(O \times a)$ on this vector will then generate a 4-dimensional surface.
\end{enumerate}

\section{Classification of embeddings with $SO(3) \ltimes T(3)$ symmetry}
Following the plan outlined in Section \ref{fullsym}, we first enumerate all possible ways to divide the embedding space into subspaces of irreducible representations of the $SO(3)$ group:
\begin{enumerate}
    \item (1, 1, 1, 1, 1, 1, 1, 1, 1, 1, 1),
    \item (1, 1, 1, 1, 1, 1, 5),
    \item (1, 1, 1, 1, 7),
    \item (1, 1, 9),
    \item (1, 1, 1, 1, 1, 1, 1, 1, 3),
    \item (1, 1, 1, 1, 1, 3, 3),
    \item (1, 1, 3, 3, 3),
    \item (1, 1, 1, 3, 5),
\end{enumerate}
where the numbers in parentheses indicate the dimensions of irreducible blocks in $V(O)$ with respect to the $SO(3)$ group.

In cases 1, 2, 3, and 4, the matrices $p_i$ turn out to be zero due to the requirement of their commutativity. In case 8, when the initial vector $\tilde y^a$ is chosen, the block corresponding to the 5-dimensional irreducible representation of $SO(3)$ becomes zero, effectively reducing this case to case 5. The remaining cases 5-7 can be considered simultaneously. Let us denote the number of scalar blocks as $M$ and the number of blocks of dimension 3 as $n$. Since $V(O \times a) \in \mathcal{P}(\mathbb{R}^{1,9})$, it must hold that $M+3n = 11$.

The three matrices $p_{i}$ must commute and consist of universal tensors. Consequently, each matrix $p_i$ consists of blocks of sizes $3\times3$, $1\times3$, $3\times1$, and $1\times1$. The $3\times3$ blocks can only be proportional to $\varepsilon_{ijk}$, the $1\times3$ and $3\times1$ blocks can only be proportional to $\delta_{ik}$, and the $1\times1$ blocks must be zero. The matrices $p_{i}$ therefore take the following form:
\begin{equation}\label{pi}
    p_{i} = \begin{pmatrix}\beta_{ab} \varepsilon_{ik_{a}j_{b}}  & \gamma_{aB} \delta_{ik_{a}} \\ \alpha_{Ab} \delta_{ij_{b}}  & 0\end{pmatrix},
\end{equation}
where it is implied that the rows of the first block are indexed by the multi-index $(a,k_a)$ and the second block by the index $A$, while the columns of the first block are indexed by $(b,j_b)$ and the second block by $B$. Here and below, indices $A,B,C = 1,...,M$, and indices $a,b,c = 1,...,n$. The commutativity condition of the matrices $p_i$ leads to the following system for the block coefficients:
\begin{equation} \label{mnConditions}
    \begin{cases}
        \beta_{ab}\beta_{bc} = \gamma_{aB} \alpha_{Bc}, \\
        \beta_{ab} \gamma_{bC} = 0, \\
        \alpha_{Ab} \beta_{bc} = 0.
    \end{cases}\
\end{equation}

In an arbitrary basis that leaves $V(O)$ in the block-diagonal form with irreducible blocks, the representation takes the form:
\begin{equation}\label{Voa}
\begin{aligned}
    &V(O \times a) = \\
    &=\begin{pmatrix}I_{ab}O_{k_{a}j_{b}} + \beta_{ab} \varepsilon_{ik_{a}m}O_{mj_{b}}a_{i} + \beta_{ac}\beta_{cb}\left( O_{l_{c}j_b}a_{k_{a}}a_{l_c} - \frac{1}{2} O_{k_{a}j_b}a^{2}\right) & \gamma_{aB} a_{k_{a}} \\ \alpha_{Ab} O_{mj_{b}} a_{m} & I_{AB}+\alpha_{Ac} \gamma_{cB} \frac{a^{2}}{2}\end{pmatrix},
\end{aligned}
\end{equation}
where the rows and columns are indexed in the same way as in \eqref{pi}, and standard summation over repeated multi-indices $(c,l_c)$ is implied. Note that there remains freedom in the choice of basis. Arbitrary changes of basis in the spaces indexed by $a,b,c$ (indices numbering three-dimensional irreducible subspaces) and by $A,B,C$ (indices numbering one-dimensional irreducible subspaces) are permissible.

The condition that the matrix $V$ has the Poincare form implies that, in some basis, the last row of the matrix $\alpha$ must be zero. Introducing a new type of indices: $\dot{A}, \dot{B}, ... = 1, ..., M-1$, there remains a freedom in the choice of basis over these indices.

A $10 \times 10$ block in the matrix $V(O \times a)$, denoted as $\Omega$ in formula \eqref{calP}, must belong to the group $SO(1,9)$. The generator of the corresponding $10\times10$ part of the matrix \eqref{Voa} is given by:
\begin{equation}\label{Aoma}
    A(\omega, a) = \begin{pmatrix}I_{ab}\omega_{k_{a}j_b} + \beta_{ab} \varepsilon_{ik_{a}j_b}a_{i} & \gamma_{a\dot{B}} a_{k_{a}} \\ \alpha_{\dot{A}b} a_{j_b}  & 0\end{pmatrix},
\end{equation}
where $\omega$ is the generator of $SO(3)$. Since $\Omega \in SO(1,9)$, certain constraints arise for such a generator $A$:
\begin{equation}\label{Aeta}
    \Omega \eta \Omega^T  = \eta \Rightarrow A \eta + \eta A^T = 0,
\end{equation}
where $\eta$ is a symmetric matrix that defines the metric of the embedding space. This equality must hold for any $\omega$ and $a_k$. In our chosen basis, $\eta$ can be split into blocks:
\begin{equation}
    \eta =
    \begin{pmatrix}
        \eta^{33}_{ak_abj_b} & \eta^{31}_{ak_a\dot{B}} \\ (\eta^{31})^T_{\dot{A}bj_b}  & \eta^{11}_{\dot{A}\dot{B}}
    \end{pmatrix}.
\end{equation}
Enforcing the equality \eqref{Aeta} for $a=0$ and arbitrary $\omega$ leads to $\eta^{31} = 0$ and $\eta^{33}_{ak_abj_b} = \tilde{\eta}^{33}_{ab}\delta_{k_aj_b}$. Consequently, the metric $\eta$ can be written as:
\begin{equation}
    \eta =
    \begin{pmatrix}
        \tilde\eta^{33}_{ab}\delta_{k_aj_b} & 0 \\ 0  & \eta^{11}_{\dot{A}\dot{B}}
    \end{pmatrix}.
\end{equation}
We can utilize the remaining freedom in the choice of basis to bring the bilinear form $\eta$ to a canonical form, where the diagonal entries can take arbitrary values of $+1$, $-1$, and $0$. Since we are seeking embeddings in $\mathbb{R}^{1,9}$, it is natural to retain only the solution $\eta = \operatorname{diag}(-1,\ldots,-1,+1)$ from all possible solutions of \eqref{Aeta}. This imposes the following conditions on the coefficients in \eqref{Aoma}:
\begin{equation}\label{conditions2}
    \begin{cases}
        \beta_{ab} = \beta_{ba},\\
        \gamma_{a \dot B} = (\alpha^T)_{a \dot C} \Tilde{\eta}_{\dot C \dot B},
    \end{cases}
\end{equation}
where $\Tilde{\eta}_{\dot C \dot B} = \operatorname{diag}(-1,\ldots,-1,+1)$. Finally, combining \eqref{mnConditions} and \eqref{conditions2} with the requirement $\alpha_{Ma} = 0$, we arrive at the final system of constraints:
\begin{equation}
    \begin{cases}
        \beta_{ab} = \beta_{ba},\\
        \gamma_{a \dot A} = (\alpha^T)_{a \dot B} \Tilde{\eta}_{\dot B \dot A}, \\
        \beta_{ac}\beta_{cb} = \gamma_{aC} \alpha_{Cb}, \\
        \beta_{ac} \gamma_{cB} = 0, \\
        \alpha_{Ac} \beta_{cb} = 0, \\
        \alpha_{Ma} = 0.
    \end{cases}
\end{equation}
It can be shown that this system is equivalent to the following simplified system:
\begin{equation}
    \begin{cases}
        \beta_{ab} = 0,\\
        \gamma_{a \dot A} = (\alpha^T)_{a \dot B} \Tilde{\eta}_{\dot B \dot A}, \\
        \gamma_{aC} \alpha_{Cb} = \gamma_{a\dot{C}} \alpha_{\dot{C}b} = 0, \\
        \alpha_{Ma} = 0.
    \end{cases}
\end{equation}

Using the obtained relations, \eqref{Voa} can now be written in the following form:
\begin{equation}
    V(O \times a) = \begin{pmatrix}I_{ab}O_{k_{a}j_b} & \gamma_{aB} a_{k_{a}} \\ \alpha_{Ab} a_{m}O_{mj_b}  & I_{AB}+\alpha_{Ac} \gamma_{cB} \frac{a^{2}}{2}\end{pmatrix}.
\end{equation}

Next, to construct the surface, it is necessary to choose the initial vector $\tilde y (t)$ so that the stability subgroup coincides with $V(SO(3))$. Considering this, specific results arise for particular values of $n,M$.
\begin{itemize}
    \item In the case $n = 3, M = 2$, no solutions can be constructed.
    \item In the case $n = 2, M = 5$, in addition to the Robertson solution, a new eight-dimensional solution can also be constructed.
    \item In the case $n = 1, M = 8$, the only possible solution is the Robertson solution.
\end{itemize}

Thus, there are only 2 symmetric embeddings of the spatially flat Friedmann model in $\mathbb{R}^{1,9}$ -- the Robertson embedding and a previously unknown eight-dimensional embedding.


\section{New embedding and dark matter}
The new 8-dimensional embedding, the construction method of which was described in the previous section, takes the form:
\begin{equation}\label{Emb8}
    \begin{aligned}
        & y^{+} = \int \frac{1}{2 \dot{f}(t)} dt+ \frac{1}{2} (x_1^{2}+x_2^2+x_3^2)f(t),\\
        & y^{-} =f(t),\\
        & y_{2} = f(t)x_1,\\
        & y_{3} = f(t)x_2,\\
        & y_{4} = f(t)x_3,\\
        & y_{5} = bx_1,\\
        & y_{6} = bx_2, \\
        & y_{7} = bx_3,
    \end{aligned}
\end{equation}
where $f(t) = \sqrt{a^{2}(t) - b^{2}}$, and $b$ is a constant. When $b \rightarrow 0$, this solution reduces to the Robertson solution.

It is straightforward to show that this embedding is unfolded. To do this, let us write out the components of its second fundamental form:
\begin{equation}
    b_{00}^{a} =\ddot{f}(t) l^{a}, \ b_{ij}^{a} = (-a \dot{a} \dot{f}) \delta_{ij} l^{a}, \ b_{i0}^{a} = \begin{pmatrix}
    x_{i}(\dot{f} - \frac{\dot{a}f}{a}) \\
    0 \\
    \delta_{ij} (\dot{f} - \frac{\dot{a}f}{a})  \\
    -\delta_{ik}b \frac{\dot{a}}{a}
    \end{pmatrix}, \ l^{a} = \begin{pmatrix}
    -\frac{1}{2 \dot{f}^{2}(t)} + \frac{1}{2} r^{2} \\
    1\\
    x_{k} \\
    0
    \end{pmatrix}.
\end{equation}
As can be seen, $b_{00}^a$ and $b_{i0}^a$ represent 4 linearly independent vectors, while the transverse space is 4-dimensional. Therefore, the surface is unfolded.

Considering the symmetry of the spatially flat Friedmann model, the tensor $\tau^{\mu\nu}$ (it appears in \eqref{bigRT} and can be interpreted as the EMT of dark matter) can be written as follows:
\begin{equation}
    \tau^{\mu}_{\nu} =
    \begin{pmatrix}
    \rho(t) & 0 & 0 & 0 \\
    0 & -p(t) & 0 & 0 \\
    0 & 0 & -p(t) & 0 \\
    0 & 0 & 0 & -p(t)
    \end{pmatrix},
\end{equation}
where the density $\rho(t)$ and pressure $p(t)$ are related by the usual equation:
\begin{equation}\label{termoD}
    \partial_0(a^3(t) \rho(t)) + (\partial_0 a^3(t)) p(t) = 0,
\end{equation}
which follows from the condition $D_\mu \tau^{\mu\nu} = 0$. From the second equation in \eqref{bigRT}, which serves as the equation of motion for dark matter in the context of this approach to its description, another condition on the functions $\rho$ and $p$ can be derived:
\begin{equation} \label{RTsuper}
    \tau^{\mu \nu} b_{\mu \nu}^{a} = \tau^{00} b_{00}^{a} + \sum\limits _{i}\tau^{ii}b_{ii}^{a} = l^{a} \left(\rho \ddot{f} - \frac{3p}{a^{2}}a \dot{a } \dot{f}\right) = 0.
\end{equation}

Together, the equations \eqref{termoD} and \eqref{RTsuper} fully determine the functions $\rho$ and $p$:
\begin{equation}
    \begin{aligned}
    &\rho =  \frac{C}{a^{3}\dot{f}}, \\
    & p = \frac{C \ddot{f}}{3 a^2 \dot{a}\dot{f}^2},
    \end{aligned}
\end{equation}
where $C$ is an integration constant, which can be arbitrarily chosen along with the constant $b$ parameterizing the embedding \eqref{Emb8}.

In the work \cite{davids01}, the possibility of describing dark matter within the framework of embedding gravity using the Robertson embedding \eqref{robertson} was proposed. A subsequent study \cite{statja26} demonstrated that within this approach, an unreasonably high density of dark matter at the beginning of inflation is required to account for the observed contribution of dark matter to the current matter density of the Universe. For the eight-dimensional embedding \eqref{Emb8} found in this work, an analysis similar to that described in \cite{statja26} was carried out. The results show that the use of this embedding also fails to provide a significant contribution of dark matter to the current matter distribution with realistic values of dark matter density at the beginning of inflation.

\section{Conclusion}
In this work, it is shown that there exist only two symmetric embeddings of the metric of the spatially flat Friedmann model into $\mathbb{R}^{1,9}$. One of these was first found by Robertson in 1933, while the second is new. In addition, we constructed two embeddings with lower symmetry -- possessing only $SO(3)$ symmetry, rather than the full symmetry of the spatially flat Friedmann model $SO(3) \ltimes T(3)$.

The new eight-dimensional symmetric embedding is unfolded, meaning that at each point, up to second-order infinitesimals, it occupies all 8 dimensions. This embedding, like the previously known Robertson embedding, can be used in attempts to explain the nature of dark matter as a fictitious matter of embedding gravity. An analysis of the equations of motion showed that the use of the new embedding does not provide any advantages compared to the Robertson embedding -- ensuring the observed amount of dark matter in the current epoch still requires assuming a high density of dark matter at the beginning of inflation.

{\bf Acknowledgements.}
The authors are grateful to A.~Sheykin for useful remarks. The work is supported by the Ministry of Science and Higher Education of the Russian Federation,
agreement no.~075-15-2022-287.


\end{document}